\documentclass[11pt]{article}
\usepackage[super,compress]{cite}
\usepackage{graphicx}
\usepackage{caption,subcaption}
\usepackage{url,hyperref}
\usepackage{color}
\usepackage{amsmath,amsfonts,amssymb}

\usepackage[figuresright]{rotating}

\newcommand{\specialcell}[2][c]{%
  \begin{tabular}[#1]{@{}c@{}}#2\end{tabular}}

\date{}
\begin{document}





\title{ANAIS: Status and prospects}
\maketitle


\begin{center}
J. Amar\'e, S. Cebri\'an, C. Cuesta, E. Garc\'ia, C. Ginestra,
M. Mart\'inez\footnote{Corresponding author. E-mail: mariam@unizar.es},
M.A. Oliv\'an, Y. Ortigoza, A. Ortiz de Sol\'orzano, C. Pobes, J. Puimed\'on, M.L. Sarsa, J.A. Villar, P. Villar
\\
\vspace{0.5cm}
Laboratorio de F\'isica Nuclear y Astropart\'iculas, Universidad de Zaragoza, Calle Pedro Cerbuna 12, 50009 Zaragoza, Spain \\
Laboratorio Subterr\'aneo de Canfranc, Paseo de los Ayerbe s/n, 22880 Canfranc Estaci\'on, Huesca, Spain
\end{center}

\begin{abstract}
ANAIS (Annual modulation with NAI Scintillators) experiment aims to look for dark matter 
annual modulation with 250~kg of ultra-pure NaI(Tl) scintillators at the Canfranc Underground Laboratory (LSC), 
in order to confirm the DAMA/LIBRA positive signal in a model-independent way. 
The detector will consists in an array of close-packed single modules, 
each of them coupled to two high efficiency Hamamatsu photomultipliers. 
Two 12.5~kg each NaI(Tl) crystals provided by Alpha Spectra are currently taking data at the LSC. 
These modules have shown an outstanding light collection efficiency (12-16 phe/keV), about the double of that from DAMA/LIBRA phase 1 detectors, which could enable reducing the energy threshold down to 1 keVee.
ANAIS crystal radiopurity goals are fulfilled for $^{232}$Th and $^{238}$U 
chains, assuming equilibrium, and in the case of $^{40}$K, present crystals activity 
(although not at the required 20~ppb level) could be acceptable. However, a $^{210}$Pb contamination 
out-of-equilibrium has been identified and its origin traced back, so we expect it will be 
avoided in next prototypes. 
Finally, current status and prospects of the experiment considering several exposure and background scenarios are presented.
\end{abstract}






\section{Introduction} \label{sec:intro}
The ANAIS experiment \cite{Amare2012, Amare2013} aims to look for dark matter (DM) annual modulation at the 
Canfranc Underground Laboratory (LSC) in order to confirm the 
DAMA/LIBRA positive DM signal \cite{bernabei2013} in a model independent way (using the same target and technique).
The design goal is to set up 250~kg of ultrapure NaI(Tl) scintillators with an energy 
threshold $\leq$2~keVee (electron equivalent units) and radioactive background near threshold below 2~c/keVee/kg/day.
In this paper we briefly report on the performances of two ANAIS prototypes that are currently taking data at LSC 
(Sec.~\ref{sec:anais25}), the ongoing activities (Sec.~\ref{sec:status}) 
and the prospects of the experiment in several exposure/background scenarios (Sec.~\ref{sec:sen}).
\section{ANAIS-25 experimental set-up} \label{sec:anais25}
The ANAIS-25 set-up, installed at LSC in December 2012 \cite{Amare2014}, consist in two 12.5~kg cylindrical 
(4.75''~$\phi~\times~$11.75'' length) NaI(Tl) crystals made by Alpha Spectra (AS), enclosed in a 30~cm lead shielding
plus anti-radon box and active vetoes. 
Each module was coupled to two Hamamatsu photomultipliers (PMTs)
through quartz windows, without light guides to improve light collection. 
The aim was to perform a comprehensive assessment of detector performances ({\it i.e.}, light collection, 
energy resolution and threshold and crystals radiopurity, in particular potassium content). On the other hand 
this set-up is being used for testing and fine-tuning the ANAIS data acquisition (DAQ) and analysis tools.

\subsection{Light collection, resolution and energy threshold}\label{sec:threshold}
\begin{figure}[h] 
\centering
\includegraphics[width=\textwidth]{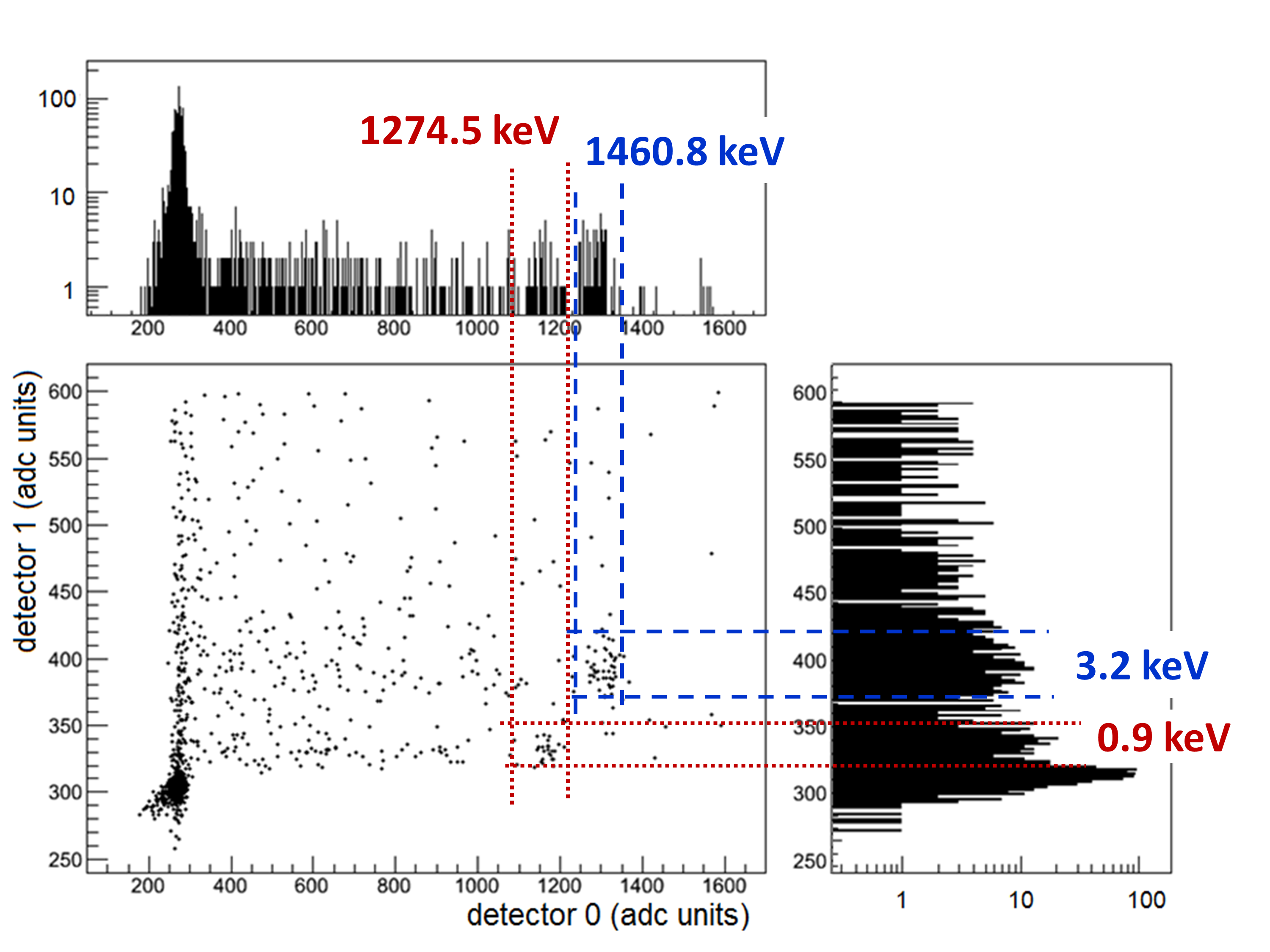}
\caption[]{Coincidence plot between both ANAIS-25 detectors. The 0.9-1274.5~keV (internal $^{22}$Na) and
3.2-1460.8~keV (internal $^{40}$K) coincident events are highlighted with dotted and dashed lines, respectively.}
\label{fig:coin}
\end{figure}

We can report an excellent light collection efficiency: 16.13$\pm$0.66~phe/keV for the 
module with high quantum efficiency PMTs (Model R12669SEL2) 
and 12.58$\pm$0.13~phe/keV for the prototype with R11065SEL PMTs.
The outstanding light collection achieved translates in a clear improvement in energy resolution and threshold
with respect to previous ANAIS prototypes. FWHM resolutions as good as 11.3 (2.8)~keV @ 122.1 (5.9)~keV have been obtained.
As regards the energy threshold, Fig.~\ref{fig:coin} shows a scatter plot of coincidence events in both detectors
during a background measurement, where the two low energy lines at 3.2 and 0.9~keV following EC in 
$^{40}$K (internal contamination) and $^{22}$Na (cosmogenic origin)
are clearly visible in one detector when the corresponding $\gamma$ line from the daughter nucleus 
de-excitation (1460.8 and 1274.5~keV respectively) is detected in the other module.
We are still working in the evaluation of the efficiency at the 0.9~keV line, but in any case a threshold of the 
order of 1~keVee seems achievable with high efficiency in the AS crystals.

\subsection{Background}\label{sec:bkg}
\begin{figure}[h] 
\centering
\includegraphics[width=\textwidth]{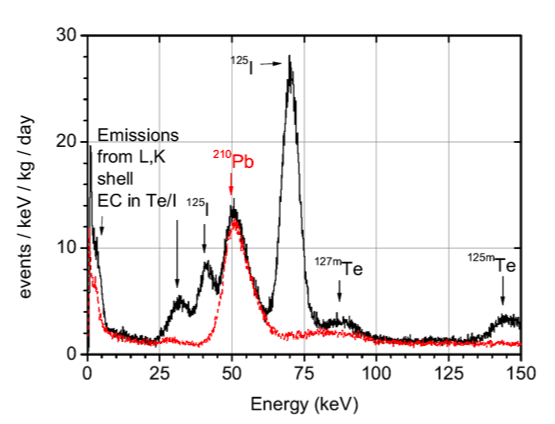}
\caption[]{Background measured in ANAIS-25 after the first month underground (black line) and
15 months after (red line). Peaks with cosmogenic origin are labeled.}
\label{fig:bkg}
\end{figure}

Fig.~\ref{fig:bkg} compares the background measured in ANAIS-25 in the month immediately following the 
mounting underground with that measured 15 months after.
We can see how the lines having a short-lived cosmogenic origin are strongly reduced,
while the non-decaying line at around 50~keV is presumably coming from internal $^{210}$Pb, as was confirmed by $\alpha$
spectroscopy.
The main crystal contaminations, found by coincidence analysis in the case of $^{40}$K \cite{Cuesta2014}, and
$\alpha$ spectroscopy plus Bi-Po sequences analysis for the radioactive chains, are given in Tab.~\ref{tab:internalCont}. 
$^{40}$K content is above ANAIS goal (20 ppb~K), but MC simulations have shown that,
for the measured contamination, 
the 3.2~keV peak can be reduced below 2~/c/keVee/kg/day by rejecting multiple hit events
thanks to the closed-packed ANAIS configuration,
obtaining an escaping probability of 25\% for the 1460.8~keV $\gamma$.
Radiopurity goals are fulfilled for $^{232}$Th and $^{238}$U chains and are sufficient for $^{40}$K, 
but a $^{210}$Pb out-of-equilibrium contamination is present at an unacceptable level, 
being responsible of the increase in background below 20~keV.
The origin of this contamination has been identified and has been solved at AS.
New material by AS could be ready very soon to be checked at LSC for radiopurity.

\begin{table}[h]
\begin{center}
\begin{tabular}{|c|c|c|c|}
\hline
$^{40}$K     &  $^{238}$U   &  $^{210}$Pb   &  $^{232}$Th \\
mBq/kg & mBq/kg & mBq/kg & mBq/kg \\
\hline
\specialcell{1.25$\pm$0.11 \\(41 ppb K)} & 0.010$\pm$0.002    &  $\sim$3.15   & 0.002$\pm$0.001 \\
\hline
\end{tabular}
\end{center}
\caption{Internal contaminations measured in ANAIS-25 prototypes. }
\label{tab:internalCont}
\end{table}

\section{ANAIS Status}\label{sec:status}
Commissioning of the experimental set-up at LSC is undergoing while radiopure enough detectors are built.
All the archaeological and low activity lead required for
the whole ANAIS shielding plus mechanical isolation, polyethylene and anti-radon box
are ready for the mounting  at the LSC.
The active vetoes that will fully cover the ANAIS set-up 
and the low energy calibration system are being tested at ANAIS-25.
Calibration at very low energy is achieved thanks to a Mylar window 
in the lateral face of the detector's copper encapsulation.
Two radioactive sources ($^{57}$Co and $^{109}$Cd) are mounted along flexible wires
that are introduced into the shielding through a closed Rn-free tube and positioned in front of the Mylar
windows.

As concerns the DAQ system, the electronic chain is fully commissioned for 
40~channels
(20 detectors x 2 PMTs), and a slow control system 
to monitorize environmental parameters (temperature at Hut/electronics/LSC Hall B, 
N$_2$ flux, HV supply, baseline noise, radon concentration at LSC Hall B, muon rate...) is almost completed.
The successive ANAIS prototypes have allowed to test the hardware as well as optimize 
the first level analysis and events selection protocols \cite{Cuesta2015}.

\section{Sensitivity prospects} \label{sec:sen}
Finally, we present the projected sensitivity of ANAIS to DM annual modulation.
If the $^{210}$Pb level is reduced in the new crystals, the main background 
contribution at low energy will come from 
internal $^{40}$K and PMTs contamination (specially $^{226}$Ra). We have considered the current 
K concentration (40~ppb) in the new crystals.
This background model (ANAIS-250) is our baseline, but we have considered also two more 
conservative scenarios in case the $^{210}$Pb suppression is not achieved:
the background level measured in ANAIS-25 and a 
background model for ANAIS-25 considering PMTs, copper encapsulation, optical
windows, lead shielding, radon in the inner volume air and NaI bulk contaminations,
based on the work presented in \cite{Cebrian2012}.
Based on the results of ANAIS-25 reported in Sec.~\ref{sec:threshold}, we 
have supposed an energy threshold of 1~keVee
and two possible total masses (100 and 250~kg) for the experiment in a 5~years data taking period.

\begin{figure}[ht] 
\centering
\includegraphics[width=\textwidth]{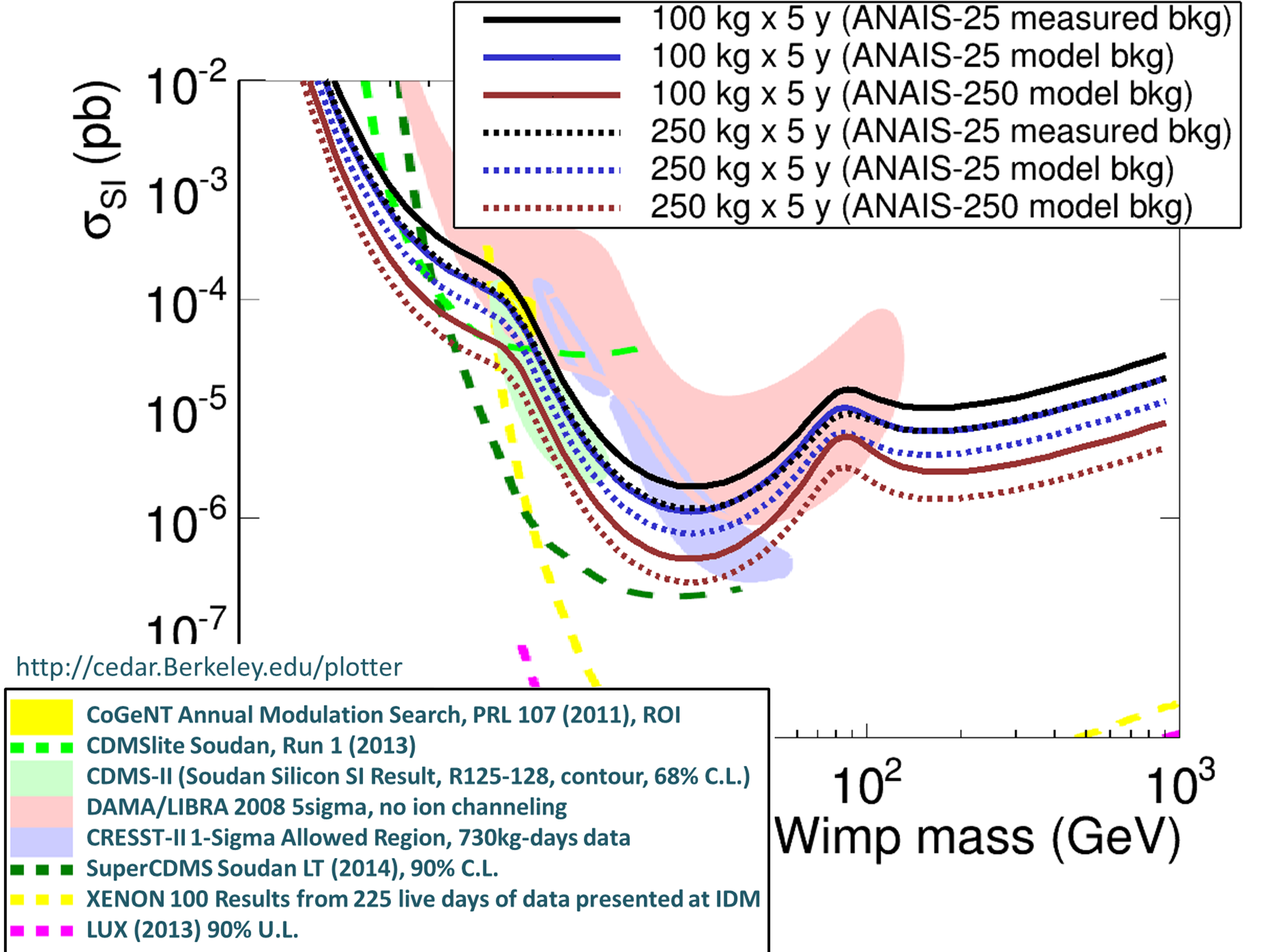}
\caption[]{Spin independent ANAIS projected sensitivity in 6 different scenarios (see text).
Comparison with DAMA-LIBRA positive result on annual modulation is model independent. }
\label{fig:sen}
\end{figure}

Fig.~\ref{fig:sen} shows the projected ANAIS spin independent sensitivity to annual
modulation supposing the three background models describe above. In the calculation, 
we have considered Helm form factors, a standard halo model (isothermal sphere) with 
$\rho_W = 0.3$~GeV/cm$^3$, v$_0 = 220$~km/s and v$_{esc} = 650$~km/s,
quenching factors 0.3 for Na and 0.1 for I, and
the energy window to look for the modulated signal is [1-3]~keVee. 
We can see how, thanks to the low energy threshold, even in the more conservative scenarios, a good sensitivity in the parameters 
region of the DAMA/LIBRA positive signal is obtained. 





\section*{Acknowledgements} 
This work has been supported by the
Spanish Ministerio de Econom\'ia y Competitividad and the
European Regional Development Fund (MINECO-FEDER)
(FPA2011-23749), the Consolider-Ingenio 2010 Programme
under grants MULTIDARK CSD2009-00064 and CPAN
CSD2007-00042, and the Gobierno de Arag\'on (Group in Nuclear
and Astroparticle Physics, ARAID Foundation and C.~Cuesta predoctoral grant). 
C.~Ginestra and P.~Villar have
been supported by the MINECO Subprograma de Formacion
de Personal Investigador. We also acknowledge LSC and
GIFNA staff for their support.
M.M. thanks the ICHEP14 organizers for a very stimulating and exciting
atmosphere.

\bibliographystyle{elsarticle-num}



\end{document}